\documentclass[12pt]{article}

        \usepackage{amsfonts}
        \usepackage{amssymb}

\usepackage{graphicx}

\def\be{\begin{eqnarray}}
\def\ee{\end{eqnarray}}
\def\bee{\begin{eqnarray*}}
\def\eee{\end{eqnarray*}}

\newtheorem{thm}{Theorem}

\newtheorem{lemma}[thm]{Lemma}

\newtheorem{prop}[thm]{Proposition}

        \def\pf{\medbreak\noindent{\bf Proof:}\enspace}
     \def\rmk{\medbreak\noindent{\bf Remark:}\enspace}
        
        \def\qed{{\bf QED}}

        \def\iff{\Leftrightarrow}
      \def\imp{\Rightarrow}
             \def\half{{\textstyle \frac{1}{2}}}
        \def\tr{\rm {Tr}\,}

\def\wh{\widehat}

\def\ds{\displaystyle}
 
\def\bra{\langle}
\def\ket{\rangle}
\def\kb{ \ket \bra }
\def\dg{\dagger}
\def\ot{\otimes}
\def\lraw{\leftrightarrow}
\def\raw{\rightarrow}

\def\half{{\textstyle \frac{1}{2}}}

\def\dtsig{{\mathbf \cdot {\bold \sigma}}}
\def\bu{{\mathbf{\bold u}}}
\def\bv{{\mathbf{\bold v}}}
\def\bw{{\mathbf{\bold w}}}
\def\bT{{\mathbf{\bold T}}}
\def\bt{{\mathbf{\bold t}}}
\def\b0{{\mathbf{\bold 0}}}
\def\rmT{{\rm T}}

          \title{Qubit Entanglement Breaking Channels}

        \author{
Mary Beth Ruskai  \thanks{Partially supported  by
 the National Security Agency (NSA) and
 Advanced Research and Development Activity (ARDA) under
Army Research Office (ARO) contract numbers
   DAAG55-98-1-0374 and DAAD19-02-1-0065, and by the National Science
        Foundation under Grant number DMS-0074566.} 
\\ Department of Mathematics \\ Tufts University\\
   Medford, Massachusetts 02155
 \\ {\normalsize ruskai@attbi.com}}

\begin{document}

\maketitle

\begin{abstract}
This paper continues the study of stochastic maps, or channels,
for which $(I \ot \Phi)(\Gamma)$ is always separable in
the case of qubits.   We give a detailed
description of entanglement-breaking qubit channels, and
show that such maps are precisely the convex hull of those
known as classical-quantum channels.
We also review the complete positivity conditions in a
canonical parameterization and show how they lead to
entanglement-breaking conditions.

\end{abstract}


 \pagebreak

\section{Introduction}

The preceding paper \cite{HRS} studied the class of stochastic maps 
which break entanglement.  For a given map $\Phi$ this
means that $I \ot \Phi(\Gamma)$
is separable for any density matrix $\Gamma$ on a tensor
product space.  It was observed that a map is entanglement
breaking if and only if it can be written in either of the
following equivalent forms
\be \label{eq:holv}
  \Phi(\rho) & = & \sum_k R_k \, \tr  F_k \rho \\
\label{eq:rank1}     & = & \sum_k 
     |\psi_k \kb \psi_k| \bra \phi_k, \rho \, \phi_k  \ket 
\ee
where each $R_k$ is a density matrix and   $ F_k $ a positive
semi-definite operator.     The map $\Phi$ is also
trace-preserving if and only if
  $\sum_k F_k = \sum_k | \phi_k \kb  \phi_k| = I$, in 
which case the set $\{ F_k \}$ form a POVM.
Henceforth we will only consider trace-preserving maps
and use the abbreviations CPT for those which are also 
completely positive and EBT for those which are also 
entanglement breaking.
An EBT map is called classical-quantum (CQ) if each
$F_k = | k \kb k |$  is a one-dimensional projection;
it is  quantum-classical (QC) if each
density matrix $R_k = | k \kb k |$ is a one-dimensional projection.

Maps which break entanglement can always be simulated 
using a classical channel; thus, one is primarily interested in
those which preserve entanglement.  Nevertheless, it is important
to understand the distinction.   In this paper we restrict
attention to EBT maps on qubits, for which one
can obtain a number of results which do not hold for general
EBT maps.  
The main new result, which does not hold in higher dimensions,
is that every qubit EBT map can be written as a convex combination
of maps in the subclass of CQ maps defined above.

Before proving this result in Section~\ref{sect:convCQ}, 
we review parameterizations
and complete positivity conditions  for qubit maps.  
We also give a number of more specialized results which use the
canonical parameterization and/or the fact that positivity of the partial 
transpose suffices to test entanglement for states on pairs of qubits.

Recall that any CPT map $\Phi$ on qubits
   can be represented by a matrix in the
canonical basis of $\{ I, \sigma_1, \sigma_2, \sigma_3 \}$.
When $\rho = \half[I + \bv \dtsig]$, then
$\Phi(\rho) = \half[ I + (\bt + \rmT \bv) \dtsig]$
where $\bt$ is the  vector with elements $t_k = t_{0k}, ~ k =1,2,3$
and $\rmT$ is a $3 \times 3$  matrix, i.e.,
${\bT} = \pmatrix{ 1 & \b0 \cr \bt & {\rmT}} $.
Moreover, it was shown in \cite{KR} that we can assume 
without loss of generality (i.e., after suitable change of bases)
 that $\rmT$ is diagonal so that $\bT$ has the canonical form
\be  \label{eq:Tcanon}
  \bT =  \pmatrix{ 1 & 0 & 0 & 0 \cr t_1 & \lambda_1 & 0 & 0 \cr
      t_2 & 0 & \lambda_2 &  0 \cr t_3 & 0 & 0 & \lambda_3 }.
\ee
The conditions for complete positivity
in this representation were obtained in \cite{RSW}
and are summarized in Section~\ref{sect:cp}.

In the case of qubits, Theorem 4  of \cite{HRS}
can be extended to give several other equivalent characterizations.
\begin{thm}  \label{thm:tfaeq} For trace-preserving qubit maps, the
 following are equivalent
\begin{itemize}
\item[A)] $\Phi$ has the Holevo form (\ref{eq:holv}) with $\{F_k \}$ a POVM.
\item[B)] $\Phi$ is entanglement breaking.
\item[C)] $\Phi \circ T $ is completely positive, where
  $T(\rho) = \rho^T$ is the transpose.
\item[D)] $\Phi$ has the ``sign-change'' property that
changing any $\lambda_k \raw -\lambda_k$ in the canonical
form (\ref{eq:Tcanon}) yields another completely positive map.

\item[E)] $\Phi$ is in the convex hull of CQ maps. 

\end{itemize}
\end{thm}
Conditions (C) thru (E) are special to qubits.  Conditions
(C) and (D) use the fact \cite{Bruss,H3,H3rev,Peres}
that the PPT (positive partial transpose)
condition for separability is also sufficient in the case of
qubits.

\section{Characterizations} \label{sect:qubit.thms} 

In this section, we prove Theorem~\ref{thm:tfaeq} and provide some
results using the canonical parameters.  This gives 
another characterization of qubit EBT maps in the special 
case of CPT maps which are also unital.

The equivalence  $(A) \iff (B)$  was proved in \cite{HRS}
where it was also shown that both are equivalent to
the condition that  $\Upsilon  \circ  \Phi$ is 
CPT for all $\Upsilon $ in a set of entanglement witnesses 
and that  $\Phi  \circ  \Upsilon$ is CPT if and only if
$\Upsilon  \circ  \Phi$ is. 
In the case of qubits, it is well-known  that it suffices
to let  $\Upsilon $ be the transpose, which proves the equivalence 
with (C). 
Furthermore, changing $\Phi \raw \Phi \circ T $
is equivalent to changing $\lambda_2 \raw -\lambda_2$ in the
representation (\ref{eq:Tcanon}), and is unitarily equivalent
(via conjugation with a Pauli matrix) to changing the
sign of any other  $\lambda_k$ which yields $(C) \iff (D)$. 
That $(E) \imp (A)$ follows immediately
from the facts that CQ maps are a special type
of entanglement-breaking maps and
the set of entanglement-breaking maps is convex by
Theorem~2 of \cite{HRS}.  The proof that
shows $(D) \imp (E)$ will be given in Section~\ref{sect:convCQ}.
~~~\qed.

The proof that (B) $\imp$ (A) given in \cite{HRS}
relied on the fact that there is a one-to-one correspondence
\cite{Choi,H3rev,J}
(but not a unitary equivalence) between maps $\Phi$ and
states 
\be \label{eq:choi}
  \Gamma_{\Phi} = (I \ot \Phi)(| \beta \kb \beta |) 
\ee
 where $| \beta \ket = \frac{1}{\sqrt{2}}( |00 \ket + |11 \ket )$ is one of
the maximally entangled Bell states.  Moreover, a map is EBT if and only
if $\Gamma_{\Phi}$ is separable since that was shown to be
equivalent to writing it in the form (\ref{eq:rank1}).  One could
then apply the  reduction criterion for separability  
\cite{BDSW,H.MP.red,H3rev} to 
$\Gamma_{\Phi}$.  This condition states that a necessary
condition for separability of $\rho$ is  that 
$ \bra \beta , \rho \, \beta \ket \leq \frac{1}{d} $ for all
maximally entangled states. 
 In the case of qubits, this criterion is equivalent to the PPT condition,
and hence sufficient, and equivalent to $\rho \leq \half I$,
which gives the following result.
\begin{thm}
A qubit CPT map is EBT if and only if
$\Gamma_{\Phi} \leq \half I$ with $\Gamma_{\Phi}$ as in 
{\em (\ref{eq:choi}).} 
\end{thm}

We now consider entanglement breaking conditions which
involve only the parameters $\lambda_k$. 
\begin{thm} \label{thm:sumleq1}
If $\Phi$ is an  entanglement breaking qubit map
written in the  form  (\ref{eq:Tcanon}),
then $\sum_j |\lambda_j| \leq 1$.
\end{thm}
\pf It is shown in \cite{AF,RSW} that a necessary condition
for complete positivity is
 \be \label{eq:cp.t0} 
(\lambda_1 \pm \lambda_2)^2 \leq (1 \pm \lambda_3)^2.
\ee
When  combined  with the sign change condition (D), this
yields the requirement
$|\lambda_1| + |\lambda_2| \leq 1 - |\lambda_3|$.  ~~\qed

For unital qubit channels, the condition in Theorem~\ref{thm:sumleq1}
is also sufficient for entanglement breaking.
For unital maps $\bt = \b0$ and, as observed in \cite{AF,KR,RSW},  the
conditions in (\ref{eq:cp.t0}) are also sufficient for complete positivity.
Since $\sum_j |\lambda_j| \leq 1$ implies that (\ref{eq:cp.t0}) holds for
any choice of sign in $\lambda_k = \pm |\lambda_k|$, it follows that
any unital CPT map satisfying this condition is also EBT.
\begin{thm} \label{thm:unital.iff}
A unital qubit channel is entanglement breaking if and only if
\linebreak $\sum_j |\lambda_j| \leq 1$ {\em [after reduction to the form
(\ref{eq:Tcanon})]}.
\end{thm}
Moreover, as will be discussed in section~\ref{sect:geom}
the extreme points of the set of unital entanglement breaking maps
are those for which two $\lambda_k = 0$.  Hence these
channels are in the convex hull of CQ maps.

For non-unital maps these conditions need not be sufficient.
Consider the so-called amplitude damping channel for which
$\lambda_1 = \alpha, \lambda_2= \alpha, \lambda_3 = \alpha^2,
t_1 = t_2 = 0$, and $t_3 = 1- \alpha^2$.   For this map equality
holds in the necessary and sufficient conditions
\be \label{eq:cp.t3ne0} 
(\lambda_1 \pm \lambda_2)^2 \leq (1 \pm \lambda_3)^2 - t_3^2 .
\ee
Since the inequalities would be violated if the sign of one
$\lambda_k$ is changed, the amplitude damping maps are never
entanglement breaking except for the limiting case $\alpha = 0$.
Thus there are maps for which 
$\sum_j |\lambda_j| = 2 \alpha + \alpha^2$ can be made arbitrarily
small (by taking $\alpha \raw 0$), but are 
{\em not} entanglement-breaking.


\section{A product representation} \label{sect:qreps}

We begin
by considering the representation of maps in
the basis $\{ I, \sigma_1, \sigma_2, \sigma_3 \}$.
  Let $\Phi$ have the form (\ref{eq:holv}) and
write $R_k = \half[I + \bw^k \dtsig]$  and
$F_k = \half[u_0^k + \bu^k \dtsig]$.
Let $W, U$ be the $n \times 4$ matrices
whose rows are
$(1, w_1^k, w_2^k, w_3^k)$ and
$(u_0^k, u_1^k, u_2^k, u_3^k)$ respectively, i.e.,
$w_{jk} = w_j^k,~u_{jk} = u_j^k ~~ k = 0 \ldots 3$.  Let ${\bT}$ be
the matrix  $W^T U$.  Note that the requirement that $\{F_k\}$  is a
POVM is precisely that the first row of $\bT$ is $(1, 0, 0, 0)$.
The matrix $\bT = W^T U$ is the representative of $\Phi$ in the
  form (\ref{eq:Tcanon}) (albeit not necessarily diagonal).
We can summarize this discussion in the following theorem.
\begin{thm} \label{thm.holv.canon}
A qubit channel is entanglement breaking if and only if it can
be represented in the form {\em (\ref{eq:Tcanon})} with $\bT = W^T U$
where $W$ and $U$ are $ n \times 4$ matrices as above, i.e.,
the rows satisfy $\big( \sum_{k=1}^3 u_{jk}^2 \big)^{1/2} \leq u_{0k}$
and 
$\big( \sum_{k=1}^3 w_{jk}^2 \big)^{1/2} \leq  w_{0k} = 1$
 for all $k$.
\end{thm}
We can use this representation to give alternate proofs
of two results of the previous section.

To show that $(A) \iff (D)$  observe that
changing the sign of the j-th column of $U$
 $(j=1,2,3)$  is  equivalent to replacing $F_k$
 by the POVM with  $u_j^k \raw -u_j^k$.  The effect on
$\bT$ is simply to multiply the $j$-th column by $-1$.
The critical property about qubits is that the condition
  $F_k > 0$ is equivalent to $(\sum_j |u_j^k|^2 )^{1/2} \leq  u_0^k $
which is unaffected by the replacement $u_j^k \raw -u_j^k$.

Next, we give an alternate proof of Theorem~\ref{thm:sumleq1} which is
of interest because it may be extendable to higher dimensions.
\pf  Let $W, U$ be as in  Section \ref{sect:qreps}.  Then
\bee
\sum_{j = 1}^3 |\lambda_j| & = &
    \sum_{j = 1}^3 \left| \sum_{k = 1}^n  w_j^k u_j^k \right| \\
    & \leq &  \sum_{j = 1}^3 \sum_{k = 1}^n  |w_j^k u_j^k  |
      = \sum_{k = 1}^n  \sum_{j = 1}^3  |w_j^k u_j^k  |  \\
   & \leq &  \sum_{k = 1}^n  \Big( \sum_{j = 1}^3  |w_j^k|^2 \Big)^{1/2}
        \Big( \sum_{j = 1}^3  |u_j^k|^2 \Big)^{1/2} \\
   & \leq  & \sum_{k = 1}^n 1 \cdot u_0^k = 1
\eee
where we have used the fact that $|\bw^k| \leq 1$ and $|\bu^k| \leq u_0^k$.
That $\sum_k u_0^k = 1$ is a consequence of the fact that the $\{ F_k \}$
form a POVM. ~~\qed

We now consider the decomposition $\bT = W^T U  $
for the special cases of CQ,  QC and point channels.
 If $\Phi$ is a CQ channel, we can assume without loss of
generality that
$ U = \half \pmatrix{ 1 & 0 & 0 & 1 \cr  1 & 0 & 0 & -1 }$.
Now write  $ W = \pmatrix{ 1 & \bw^1 \cr  1 & \bw^2 }$.
Then
\be
  \bT = W^T U = \pmatrix{ 1 & 0 & 0 & 0 \cr ~ & 0 & 0 & \cdot \cr
   \frac{ \bw^1 + \bw^2}{2} & 0 & 0 & \frac{ \bw^1 - \bw^2}{2}
    \cr  ~ & 0 & 0 & \cdot }.
\ee
By acting on the left with a unitary matrix of the form
$\pmatrix{ 1 & \b0 \cr \b0 & \pm{\cal R}}$ where  ${\cal R}$
is a rotation whose third row is a multiple of $\bw^1 - \bw^2$,
this can be reduced to the form (\ref{eq:Tcanon}) with
$ \lambda_1 = \lambda_2 = 0$, $ |\lambda_3| =  \half | \bw^1 - \bw^2 |$,
and $\bt = {\cal R} \bw^1 - \pmatrix{0, 0, \lambda_3}^T$
[since  $\half(\bw^1 + \bw^2) = \bw^1 - \half(\bw^1 - \bw^2)$.]
Indeed, it suffices to choose
\be  \label{eq:CQmatW}
 W =  \pmatrix{ 1 & t_1 & t_2 & t_3 + \lambda_3 \cr
   1 & t_1 & t_2 & t_3 - \lambda_3 }.
\ee
  Note that  the requirement
$|\bt| \leq 1$ only implies $t_1^2 + t_2^2 + (t_3 + \lambda_3)^2 \leq 1$;
however, the requirement $|\bw^k| \leq 1$ implies that
$t_1^2 + t_2^2 + (t_3 \pm \lambda_3)^2 \leq 1$ must hold with both
signs and this is equivalent to the stronger condition
\be  \label{eq:nasc.lin}
  t_1^2 + t_2^2 + (|t_3| + |\lambda_3|)^2 \leq 1
\ee
which is necessary and sufficient for a CPT map to reduce
the Bloch sphere to a line.

If $\Phi$ is a QC channel, we can assume without loss of generality that
$ W =  \pmatrix{ 1 & 0 & 0 & 1 \cr  1 & 0 & 0 & -1 }$
and $ U =
  \pmatrix{ u_0 & u_1 & u_2 & u_3 \cr  1- u_0 & -u_1 & -u_2 & -u_3 }$,
from which one easily finds that the second and third rows of
$\bT = W^T U$ are identically zero and the fourth row is
$\pmatrix{ 2u_0 -1 & 2 u_1 & 2 u_2 & 2 u_3 }$.  One then easily
verifies that multiplication on the right by a matrix as above
with  ${\cal R}$ a rotation whose third column is a multiple
of $\pmatrix{u_1 & u_2 & u_3 }$ reduces $\bT = W^T U$ to the
canonical form (\ref{eq:Tcanon}) with
$ \lambda_1 = \lambda_2 = 0$,
$ \lambda_3 = 2 \sqrt{u_1^2 + u_2^2  + u_2^2} = |\bu| \leq
 \min \{ 2u_0, 2(1-u_0) \} \leq 1$, and   $t_3 = 2 u_0 - 1$.
(Note that $t_3 +  \lambda_3 \leq |2 u_0 - 1| +
     \min \{ 2u_0, 2(1-u_0) \} \leq 1$ with equality if and
only if the image reaches the Bloch sphere.)

It is interesting to note that for qubits channels, every QC channel
is unitarily equivalent to a CQ channel.  Indeed, a channel which,
after reduction to canonical form has non-zero elements
$\lambda_3$ and $t_3$ with $|\lambda_3| + |t_3| \leq 1$ and
$|t_3| < 1$ can
be written as either a QC channel with
\bee
  W =  \pmatrix{ 1 & 0 & 0 & 1 \cr  1 & 0 & 0 & -1 }
~~  U =  \half \pmatrix{ 1 + t_3 & 0 & 0 & \lambda_3 \cr
     1 - t_3 & 0 & 0 & -\lambda_3 }
\eee
or as a CQ channel with
\bee
  W =  \pmatrix{ 1 & 0 & 0 & t_3 + \lambda_3 \cr  1 & 0 & 0 &
t_3 - \lambda_3  }   ~~
  U =  \half \pmatrix{ 1 & 0 & 0 & 1 \cr  1 & 0 & 0 & -1 } .
\eee
For point channels $W = \pmatrix{ 1 & t_1 & t_2 & t_3}$
and $U =  \half \pmatrix{ 1 & 0 & 0 & 0}$.


We conclude this section with an example of map 
of the form (\ref{eq:holv}) with an extreme POVM,
 for which the corresponding map $\Phi$ is {\em not} extreme.
Let  $E_k = \frac{1}{3}[I + \bw^k \dtsig]$ with
$\bw^1 = (1, 0, 0), \bw^2 = (-\half, 0 ,\frac{\sqrt{3}}{2}),
  \bw^3 = (-\half, 0 ,-\frac{\sqrt{3}}{2})$.
Then, irrespective of the choice of $R_k$, the third column
of $\bT = W^T U$ is identically zero, which implies that,
after reduction to canonical form, one of the parameters
$\lambda_k = 0$.  However, it is easy to find density matrices, e.g.,
$R_k = \half[I + \sigma_k]$, for which the resulting map $\Phi$ is not
CQ or point.  But by Theorem~\ref{thm:CQqubit},  $\Phi$ is a convex
combination of CQ maps and hence, not extreme.

\section{Complete positivity conditions revisited} \label{sect:cp}

Not only is the set of CPT maps convex, in a fixed
basis corresponding to the canonical form (\ref{eq:Tcanon})
the set of $\lambda_k$ corresponding to any fixed choice
of $\bt = (t_1,t_2,t_3)$ is also a convex set which we denote
$\Lambda_{\bt}$.  We will also be interested in the convex subset
$\Lambda_{\bt, \lambda_3}$ of the $\lambda_1$-$\lambda_2$ plane
for fixed $\bt, \lambda_3$, and in the convex set
$\Xi_{t_3, \lambda_3}$ of points $(t_1,t_2, \lambda_1, \lambda_2)$
corresponding to fixed $t_3, \lambda_3$.
Although stated somewhat differently, the following result was proved in
\cite{RSW}.
\begin{thm} \label{thm:RSW}
Let $\bt$ and $\lambda_3$ be fixed with $|t_3| + |\lambda_3| < 1$.
Then the convex set $\Lambda_{\bt, \lambda_3}$ consists of the
points $(\lambda_1,\lambda_2)$ for which
$I - R_{\Phi}^{\dg} R_{\Phi}$ {\em (or, equivalently
$I - R_{\Phi}R_{\Phi}^{\dg}$)} is positive  semi-definite,
where
\be \label{eq:cont.mtrx}
R_{\Phi} = \pmatrix{  \frac{t_1+it_2} {(1+t_3+\lambda_3)^{1/2}
   (1-t_3-\lambda_3)^{1/2}} &
 \frac{\lambda_1+\lambda_2}{(1+t_3+\lambda_3)^{1/2}
      (1-t_3+\lambda_3)^{1/2}}      
\cr ~~& ~~  \cr
\frac{\lambda_1-\lambda_2} {(1+t_3-\lambda_3)^{1/2}
(1-t_3-\lambda_3)^{1/2}} &
 \frac{t_1+it_2}{(1+t_3-\lambda_3)^{1/2}
(1-t_3+\lambda_3)^{1/2}  }}.
\ee
Similarly,  $\Xi_{t_3, \lambda_3}$ also consists of the points
$(t_1,t_2, \lambda_1, \lambda_2)$ for which
$I - R_{\Phi}^{\dg} R_{\Phi} \geq 0$.  Moreover, the extreme
points of $\Lambda_{t_3, \lambda_3}$ are those for which
$R_{\Phi}^{\dg} R_{\Phi} = I$.
\end{thm}
Although this result is stated in a form in which $t_3$ and $\lambda_3$
play a special role and does not appear to be symmetric with respect
to interchange of indices, the conditions which result are, in fact,
invariant under permutations of $1,2,3$.

Theorem~\ref{thm:RSW} follows from Choi's  theorem \cite{Choi} that
$\Phi$ is completely positive if and only if 
$\Gamma_{\Phi}$,    given by (\ref{eq:choi}), is
positive semi-definite.
As noted in \cite{RSW}, this implies that it can be written in the form
\be
 \Gamma_{\Phi} = \pmatrix{
   \Phi(E_{11}) & \sqrt{\Phi(E_{11})} R_\Phi  \sqrt{\Phi(E_{22})} \cr
\sqrt{\Phi(E_{22})} R_\Phi^{\dg}  \sqrt{\Phi(E_{11})} & \Phi(E_{22}) }
\ee
where $R_\Phi $ is a contraction.  (Note, however, that the 
expression for $R_\Phi $ given in (\ref{eq:cont.mtrx}) was
obtained by applying  this result to the adjoint $\wh{\Phi}$,
i.e, to $(I \ot \wh{\Phi})(| \beta \kb \beta |)$. 

Conversely,
 given a CPT map $\Phi$ and {\em any} contraction $U$ on
${\bf C}^2$, one can define a $4 \times 4$ matrix in block form,
\be  \label{eq:genU}
M =   \pmatrix{ \wh{\Phi}(E_{11})  &
  \sqrt{\wh{\Phi}(E_{11})} \, U \sqrt{\wh{\Phi}(E_{22})} \cr
 \sqrt{\wh{\Phi}(E_{22})} \, U^{\dg} \sqrt{\wh{\Phi}(E_{11})}  &
    \wh{\Phi}(E_{22})}
\ee
It  then follows that there is another CPT map which
(with a slight abuse of notation) we denote
$\Phi_{U}$ for which $(I \ot \wh{\Phi_{U}})(| \beta \kb \beta |) = M$.
However, (\ref{eq:genU}) need not, in general,
correspond to a map $\Phi_{U}$ which has the canonical
form (\ref{eq:Tcanon}) since that requires
$\wh{\Phi}(E_{12}) = \sqrt{\wh{\Phi}(E_{11})} \, U \sqrt{\wh{\Phi}(E_{22})}
  = (t_1 + i t_2) I + \lambda_1 \sigma_x + i \lambda_2 \sigma_y$. 
For $U$ an arbitrary unitary or contraction, we can only
conclude that 
\bee
  \wh{\Phi}(\sigma_x) & = ~
   \sqrt{\wh{\Phi}(E_{11})} \, U \sqrt{\wh{\Phi}(E_{22})} +
 \sqrt{\wh{\Phi}(E_{22})}  \, U^{\dg} \sqrt{\wh{\Phi}(E_{22})}
  & \equiv ~\sum_{k=0}^3 t_{1k} \, \sigma_k \\
  \wh{\Phi}(\sigma_y) & = ~
  \sqrt{\wh{\Phi}(E_{11})} \, U \sqrt{\wh{\Phi}(E_{22})} -
  \sqrt{\wh{\Phi}(E_{22})} \, U^{\dg} \sqrt{\wh{\Phi}(E_{22})} 
&  \equiv  ~ \sum_{k=0}^3 t_{2k} \, \sigma_k
\eee
so that
the map $\Phi_{U}$ corresponds to a matrix of the form 
  $ \pmatrix{ 1 & 0 & 0 & 0 \cr t_{10} & t_{11} & t_{12} & t_{13} \cr
      t_{20} & t_{21} & t_{22} & t_{23} \cr t_3 & 0 & 0 & \lambda_3 }$
with $t_{jk}$ real.


In order to study the general case of non-zero $t_k$, it is convenient to
rewrite (\ref{eq:cont.mtrx}) in the following form  (using
 notation similar to that introduced in \cite{King1}).
\be
  R_{\Phi} = \pmatrix{ \ds{\frac{\tau}{\sqrt{\, c_{++} \, c_{--}}}} &
  \ds{\frac{\lambda_+}{\sqrt{\, c_{++} \, c_{+-}}}}  \cr
  \ds{\frac{\lambda_-}{\sqrt{\, c_{--} \, c_{-+}}}}  &
 \ds{\frac{\tau}{\sqrt{\, c_{+-} \, c_{-+}}}} }
\ee
where $\lambda_{\pm} = \lambda_1 \pm \lambda_2$,
 $\tau = t_1 + i t_2$, and $c_{\pm \pm} = 1 \pm \lambda_3 \pm t_3$,
e.g., $c_{+-} = 1 + \lambda_3 - t_3$.
Then
\be
  I - R_{\Phi}^{\dg} R_{\Phi} \equiv M =
  \pmatrix{ m_{11} & m_{12} \cr m_{21} & m_{22} }
\ee
with
\be
   m_{11} & = &  1 - \frac{|\tau|^2}{c_{++} \, c_{--} } -
      \frac{|\lambda_-|^2}{c_{--} \, c_{-+} }
    \label{eq:diag1} \\   & ~ & \nonumber \\
m_{22} & = &  1 - \frac{|\tau|^2}{c_{+-} \, c_{-+} } -
      \frac{|\lambda_+|^2}{c_{++} \, c_{+-} }
    \label{eq:diag2} \\ & ~ &  \nonumber \\
m_{12} = \overline{m}_{21} & = &
  \frac{\overline{\tau} ~ \lambda_+}{ c_{++} \sqrt{\, c_{--} \, c_{+-}}} +
    \frac{\tau ~ \lambda_-}{ c_{-+} \sqrt{\, c_{--} \, c_{+-}}}
   \label{eq:offdiag}
\ee
Note that the denominators, although somewhat messy, are essentially
constants depending only on $t_3$ and $\lambda_3$.  Considering
$\tau$ as also a fixed constant it suffices to rotate (and dilate) the
$\lambda_1$-$\lambda_2$ plane by $\pi/4$ and work instead with the
variables $\lambda_{\pm}$.  

\begin{figure}
\hskip1cm
\includegraphics*[width=15cm,height=15cm,keepaspectratio=true]{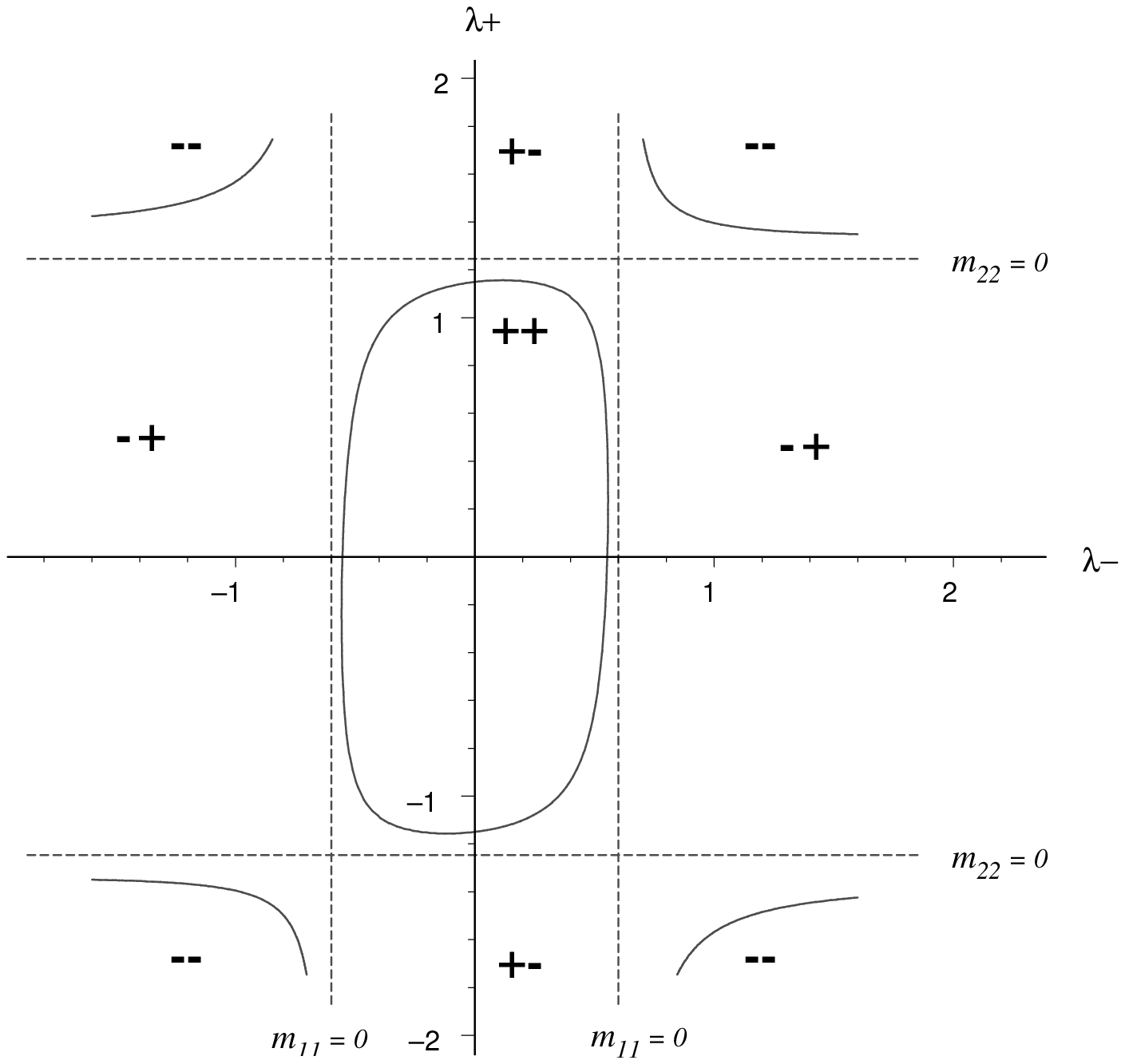}
\caption{The $\lambda_+$-$\lambda_-$ plane showing the
regions described by the diagonal conditions (dotted lines)
and the curves corresponding to 
$\det(I - R_{\Phi}^{\dg} R_{\Phi}) = 0$
for $\bt = (0.2,0.3,0)$ and $\lambda_3 = 0.35$.
The closed curve and its interior describes the parameters
for which the corresponding map is completely positive.}
\label{fig:regions}
\end{figure}

The diagonal conditions  $m_{11} \geq 0$ and $m_{22} \geq 0$
define a rectangle in the $\lambda_+$-$\lambda_-$ plane, namely
\be
  |\lambda_-|^2 & \leq & c_{--} \, c_{-+} - \frac{c_{-+}}{c_{++}} \,
|\tau|^2
  =  (1- \lambda_3)^2 - t_3^2 -
      \frac{1 - \lambda_3 + t_3}{1 + \lambda_3 + t_3} \, |\tau|^2
 \label{eq:m11>0}   \\  & ~ & \nonumber \\  \label{eq:m22>0}
  |\lambda_+|^2 & \leq & c_{++} \, c_{+-} - \frac{c_{++}}{c_{-+}}\, |\tau|^2
  =  (1+ \lambda_3)^2 - t_3^2 -
      \frac{1+ \lambda_3 + t_3}{1 - \lambda_3 + t_3} \, |\tau|^2
\ee
These diagonal conditions imply the necessary conditions
\be  \label{eq:cp.tau0}
  |\lambda_{\pm}|^2 \leq (1 \pm \lambda_3)^2 - t_3^2
\ee
for complete positivity, which also become sufficient when $\tau = 0$.
The determinant condition $m_{11} m_{22} \geq |m_{12}|^2$ is more
complicated, but basically has the form
\be   \label{eq.detcond}
  \big[a-b \lambda_+^2 \big] \, \big[c - d \lambda_-^2 \big] \geq
       e \lambda_+^2 + f \lambda_-^2 + g \lambda_+ \, \lambda_-
\ee
  In particular, we would like
to know if the values of $(\lambda_+, \lambda_-)$ satisfying
(\ref{eq.detcond}) necessarily lie within the rectangle defined by
(\ref{eq:m11>0}) and (\ref{eq:m22>0}) .
Extending the lines bounding this rectangle, i.e.,
$m_{11} = 0$ and $ m_{22} =  0$ one sees that the
$\lambda_+$-$\lambda_-$ plane is divided into 9 regions, as
shown in Figure~\ref{fig:regions} and described below.
\begin{itemize}
\item the rectangle in the center which we denote $++$,
\item four (4)  outer corners which we denote $--$ since both
$m_{11} < 0$ and $ m_{22} < 0$,
\item the four (4) remaining regions (directly above, below and to the
left and right of the center rectangle) which we denote as $+-$ or $-+$
according to the signs of $m_{11}$ and $ m_{22}$.
\end{itemize}
We know that the determinant condition (\ref{eq.detcond}) is
{\em never} satisfied in the $+-$ or $-+$ regions since
$m_{11} m_{22} - |m_{12}|^2 < 0$ when $m_{11}$ and $ m_{22}$ have
opposite signs.  This implies that equality in (\ref{eq.detcond})
defines a curve which bounds a convex region lying entirely
within the $++$ rectangle.  Although (\ref{eq.detcond}) also has
solutions in the $--$ regions as shown in Figure~\ref{fig:regions}, one
expects that these will typically lie {\em outside} the region
for which $|t_k|+ |\lambda_k| \leq 1$, i.e., the rectangle
bounded by the line segments satisfying
$|\lambda_{+} + \lambda_{-}| \leq 2( 1- |t_1|)$
and $|\lambda_{+} - \lambda_{-}| \leq 2( 1- |t_2|)$.
However, John Cortese \cite{Cort} has shown that this need not
necessarily  be the case.
Nevertheless, one need only  check one of the two conditions 
$m_{11} > 0$,
$ m_{22} > 0$, and might substitute a weaker condition, such as
$\tr M > 0$, to exclude points in the $--$ regions.
For example, one could substitute for the diagonal
conditions, 
$ c_{--}  m_{11} +  c_{+-} m_{22} \geq 0 $
which is equivalent to 
\be   \label{eq:diag.alt}
  \big(\lambda_1^2 + \lambda_2^2 \big)(1 + t_3) + 
  \lambda_3^2 (1 - t_3) \leq (1 + t_3) (1 - |\bt|)^2
   + 2 \lambda_1 \lambda_2 \lambda_3 .
\ee
Thus, strict inequality in both (\ref{eq.detcond}) and
(\ref{eq:diag.alt}) suffice to ensure complete positivity.

\begin{figure}
\hskip3cm
\includegraphics*[width=8cm,height=8cm,keepaspectratio=true]{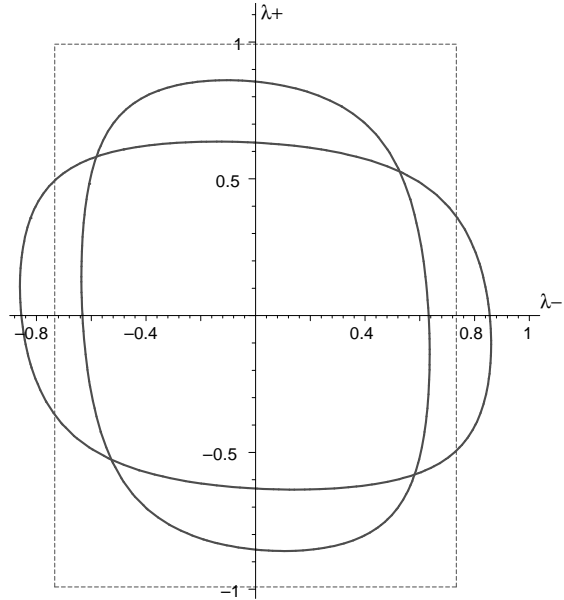}
\caption{The $\lambda_+$-$\lambda_-$ plane showing 
the region determined by determinant condition 
when $\bt = (0.4,0.3,0.0)$ and $\lambda_3 = 0.15$ and
the corresponding region with $\lambda_+$ and$\lambda_-$
interchanged.  Their intersection corresponds to the
entanglement breaking maps with the indicated parameters.}
\label{fig:swap}
\end{figure}

In general, when $\bt \neq 0$, the convex set 
$\Lambda_{\bt,\lambda_3}$ is determined by (\ref{eq.detcond}),
i.e, by the closed  curve for which equality holds and its 
interior.  Since changing the sign of
$\lambda_1$ or $\lambda_2$ is equivalent to changing
 $\lambda_+ \lraw\lambda_{-}$, the corresponding
set of entanglement breaking maps is given by the 
intersection of this region with the corresponding one
with $\lambda_+$ and $\lambda_{-}$ switched, as shown in 
Figure~\ref{fig:swap}.

  \smallskip

\rmk If, instead of looking at $I - R_{\Phi}^{\dg} R_{\Phi}$, we
had considered $I - R_{\Phi} R_{\Phi}^{\dg}$, the matrix $M$
would change slightly and the conditions
(\ref{eq:m11>0}) or (\ref{eq:m22>0}) would be modified accordingly.
(In fact, the only change would be to replace $+t_3$ by $-t_3$
in the fraction multiplying $|\tau|^2$.)  However, the
determinant condition (\ref{eq.detcond}) would {\em not} change.
Since $R_{\Phi}^{\dg} R_{\Phi}$ and $R_{\Phi} R_{\Phi}^{\dg}$, are
unitarily equivalent,
$$ \det[I - R_{\Phi}^{\dg} R_{\Phi}] =
   \det \Big( U[I - R_{\Phi} R_{\Phi}^{\dg}] U^{\dg} \Big)
   =  \det[I - R_{\Phi} R_{\Phi}^{\dg}]. $$


It is worth noting that  whether or not $R_{\Phi}$
   is a contraction is {\em not} affected by the
signs of the $t_k$.  (In particular, changing $t_2 \mapsto -t_2$
 takes  $R_{\Phi} \mapsto \overline{R}_{\Phi}$,  
changing  $t_3 \mapsto -t_3$ takes 
$R_{\Phi} \mapsto \sigma_x R_{\Phi}^T \sigma_x$, and 
changing  $t_1 \mapsto -t_1$ takes 
$R_{\Phi} \mapsto -\sigma_z \overline{R}_{\Phi} \sigma_z$.)
Therefore, one can change the sign of any {\em one} of the $t_k$ 
without affecting completely positivity.   

By contrast, one can {\em not}, in general,
 change $\lambda_k \raw - \lambda_k$ without affecting the complete
positivity conditions. (Note, however, that one can always change the
signs of any {\em two} of the
$\lambda_k$ since this is equivalent to conjugation
with a Pauli matrix on either the domain or range.  The latter
will also change the signs of two of the $t_k$.)
Changing the sign of $\lambda_2$ is equivalent to composing $\Phi$
with the transpose, so that changing the sign of one of
the $\lambda_k$ is equivalent to composing $\Phi$ with the
transpose and conjugation with one of the Pauli matrices.
Furthermore, if changing the sign of one particular $\lambda_k$ does
not affect complete positivity, then one can change the sign of
{\em any} of the $\lambda_k$ without affecting complete positivity.

In view of the role of the sign change condition it is
worth summarizing these remarks.

    
\begin{prop}
Let $\Phi$ be a CPT map in canonical form (\ref{eq:Tcanon})
and let $T(\rho) = \rho^T$ denote the transpose.  Then
\begin{itemize}
\item[{\rm (i)}] $T \circ \Phi \circ T$ is also completely positive, i.e.,
changing $t_k \raw - t_k$ does not affect complete positivity.
\item[{\rm (ii)}] $ \Phi \circ T$ is completely positive if and only
if changing any $\lambda_k \raw - \lambda_k$ does not
affect complete positivity.
\item[{\rm (iii)}] $ \Phi \circ T$ is completely positive if and only
$T \circ \Phi$ is.
\end{itemize}
\end{prop}
The only difference between  $ \Phi \circ T$ and $T \circ \Phi$
is that the former changes the sign of $\lambda_2$ while
the latter changes the signs of both $t_2$ and $\lambda_2$.


\section{Geometry} \label{sect:geom}

\noindent {\bf Image of the Bloch sphere} 

\smallskip

We first consider the geometry of entanglement breaking channels
in terms of their  effect on the Bloch sphere.
It  follows from the equivalence with the sign change condition in
Theorem~\ref{thm:tfaeq} that any CPT map with some 
$\lambda_k = 0$ is
entanglement breaking.   We call such channels {\em planar}
since the image lies in a plane within the Bloch sphere.
Similarly, we call a channel with two
$\lambda_k = 0$ {\em linear}.  If all three  $\lambda_k = 0$, the Bloch
sphere is mapped into a point.
 Note that the subsets of channels whose images
lie within points, lines, and planes respectively are {\em not} convex.
However, they are well-defined and useful classes to consider.

\smallskip

\noindent{\bf Points:}  A channel which maps
the Bloch sphere to a point has the Holevo form (\ref{eq:holv})
in which the sum  reduces to a single term with 
$R = \half[I + \bt \dtsig]$ and $E = I$.  Then
$\Phi(\rho) = R \, \tr(E \rho) = R ~\forall ~ \rho$ and
${\bf T} = \pmatrix{ 1 & \b0 \cr \bt & \b0} $.
when $|\bt| = 1$,  $R$ is a pure state and the map is extreme.
It is also a special case of the so-called amplitude
damping channels, and (as noted at the end
of section~\ref{sect:qubit.thms}) these are the only 
amplitude damping channels which break  entanglement.

\smallskip

\noindent{\bf Lines:} When two of the $\lambda_k = 0$ so that the image of
the Bloch  sphere is a line, the conditions for complete positivity reduce
to a single inequality, which becomes (\ref{eq:nasc.lin})
in the case $\lambda_1 = \lambda_2 = 0$.
Moreover, it is straightforward to verify that any such channel
can be realized as a CQ channel.
Indeed, it suffices to choose $W$ as in (\ref{eq:CQmatW}).

\smallskip

\noindent{\bf Planar channels:} The image of a map with exactly 
one $\lambda_k = 0$ lies in a plane.   When
this is $\lambda_3$, the condition
$I - R_{\Phi}^{\dg} R_{\Phi} \geq 0$ becomes
\bee
 \pmatrix{ 1 - |\bt|^2 - (\lambda_1 - \lambda_2)^2 &
   2(t_1 \lambda_1 + i t_2 \lambda_2)  \cr
   2(t_1 \lambda_1 - i t_2 \lambda_2)
   & 1 - |\bt|^2 - (\lambda_1 + \lambda_2)^2} \geq 0 .
\eee
where $|\bt|^2 = t_1^2 +  t_2^2 + t_2^2$, and the condition on
the diagonal becomes
\be \label{eq:nasc.planar}
(|\lambda_1| + | \lambda_2|)^2 + |\bt|^2 \leq 1 .
\ee
Now, if either 
diagonal element is identically zero, then one must have
$t_1 \lambda_1 = t_2 \lambda_2 = 0$.
Thus, if both
$\lambda_1, \lambda_2 \neq 0$ and equality holds in
the necessary condition (\ref{eq:nasc.planar}),
one must have $t_1 = t_2 = 0$, in which case it
reduces to $(|\lambda_1| + | \lambda_2|)^2 + t_3^2 = 1$.
This implies that a truly planar channel  can not touch
the Bloch sphere, unless it reduces to a point or a line.

  \medskip

\noindent {\bf Geometry of $\lambda_k$ space}

\smallskip

We now consider, instead of the geometry of the images of
entanglement-breaking maps, the geometry of the allowed
set of maps in $\lambda_k$ space.
After reduction to the canonical form (\ref{eq:Tcanon}) it is
often useful to look at the subset of $[\lambda_1, \lambda_2, \lambda_3]$
which correspond to a particular class of maps.  We first
consider maps for which $\bt = 0$.

\begin{thm} \label{thm:octa}
In a fixed (diagonal) basis, the set of unital entanglement breaking
maps on qubits corresponds to
 the octahedron whose extreme points correspond to the channels
for which $[\lambda_1, \lambda_2, \lambda_3]$ is a permutation of
$[\pm 1, 0, 0]$.
\end{thm}
Since this octahedron is precisely the subset with
$\sum_j |\lambda_j| \leq 1$ the result follows immediately
from Theorem \ref{thm:unital.iff}.  Alternatively, one
could use Theorem \ref{thm:CQqubit} and the fact that the 
unital CQ maps must have the form above.

\medskip

\noindent{\bf Remarks:} 
\begin{enumerate}
\item The channels
corresponding to  a permutation of $[\pm 1, 0, 0]$ belong
to the subclass known as CQ channels.   Hence, the set of unital
entanglement breaking maps is the convex hull of unital
CQ maps.

\item   This octahedron in Theorem~\ref{thm:octa} is precisely the
intersection of the tetrahedron with corners
$[1,1,1],~[1,-1,-1],~[-1,1,-1], ~[-1,-1,1]$ with its inversion through
the origin, as shown in Figure~\ref{fig:octa}. (A similar picture arises
in studies of entanglement and Bell inequalities.  See, e.g., Figure 3 in
\cite{VW} or Fig. 2 in \cite{BNT}). 

\item  The tetrahedron of unital maps is precisely the intersection
of the four planes of the form
    $ {\bf n} \cdot [\lambda_1, \lambda_2, \lambda_3] = 1$
with ${\bf n} = [\pm1, \pm1, \pm1]$ and an odd number of negative
signs, i.e., $n_1 n_2 n_3 = -1$.   The octahedron of unital EBT maps
is precisely the intersection of all eight planes of this form.

\item   If the octahedron of unital entanglement breaking maps is removed
from the tetrahedron of unital maps, one is left with four disjoint
tetrahedrons whose sides are half the length of the original.
Each of these defines a region of ``entanglement-preserving"
unital channels with fixed sign.  For example, the tetrahedron
with corners, $[1,1,1],~[1,0,0],~[0,10], ~[1,0,0]$; this   is
the interior of the intersection of the plane
$ [-1,-1,-1] \cdot [\lambda_1, \lambda_2, \lambda_3] = -1$
and the three planes of the form
    $ {\bf n} \cdot [\lambda_1, \lambda_2, \lambda_3] = 1$
with ${\bf n} = [1,1,-1],~ [1,-1,1], [-1,1,1]$.  For many purposes,
e.g., consideration of additivity questions, it suffices to
confine attention to one of these four corner tetrahedrons.
Indeed, conjugation with one of the Pauli matrices, transforms
the corner above into one of the other four.
\end{enumerate}

We next consider non-unital maps, for which
one finds the following analogue of Theorem \ref{thm:octa}.

\begin{thm}  \label{thm:sign}
Let $\bt = (t_1, t_2, t_3)$ be a fixed vector in ${\bf R}^3$
and let  $\Lambda_\bt$ denote the convex subset of ${\bf R}^3$
corresponding to the vectors $[\lambda_1, \lambda_2, \lambda_3]$
for which the canonical map with these parameters is completely
positive.   Then the intersection of $\Lambda_\bt$ with its
inversion through the origin (i.e., $\lambda_j \raw - \lambda_j$)
is the subset of EBT maps with translation $\bt$.
\end{thm}

\rmk The effect of changing the sign of $\lambda_2$ is
$\lambda_+ \lraw \lambda_-$ and of changing the sign of
$\lambda_1$ is $\lambda_+ \lraw -\lambda_-$.  In either
case, the effect on the determinant condition (\ref{eq.detcond})
is simply to switch $\lambda_+ \lraw \lambda_-$, i.e, to
reflect the boundary across the $\lambda_+ = \lambda_-$ line.
Thus, the intersection of these two regions will correspond to
entanglement breaking channels.  The remainder will, typically,
consist of 4 disjoint (non-convex) regions, corresponding to the
four corners remaining after the ``rounded octahedron'' of
Theorem \ref{thm:sign} is removed from the ``rounded tetrahedron''.



\section{Convex hull of qubit CQ maps}   \label{sect:convCQ}

In \cite{RSW} we found it useful to generalize  the extreme points
of the set of CPT maps ${\cal S}$ to include all
 maps for which $R_{\Phi}$ is unitary, which is equivalent
to the statement that {\em both} singular values of $R_{\Phi}$ are $1$.
In addition to true extreme points, this includes
``quasi-extreme'' points which correspond to the edges of
the tetrahedron of unital maps.   Some of these quasi-extreme
points are true extreme points for the set of entanglement-breaking
maps.  However, there are no extreme points of the latter which
are not generalized extreme points of ${\cal S}$.  This will
allow us to conclude the following.
\begin{thm}  \label{thm:CQqubit}
Every extreme point of the set of entanglement-breaking qubit
maps is a CQ map.  Hence, the set of  entanglement-breaking qubit
maps is the convex hull of qubit CQ maps.
\end{thm}
The goal of the section is to prove this result.   Because
our argument is somewhat subtle, we also include, at the
end of this section a direct proof of some special cases.

First we note that the following was shown in \cite{RSW}.
After reduction to canonical form (\ref{eq:Tcanon}), for
any map which is a generalized extreme point, the parameters
$\lambda_k$ must satisfy (up to permutation)
$\lambda_3 = \lambda_1 \lambda_2$.   This is compatible
with the sign change condition if and only if at least
two of the $\lambda_k = 0$, which implies that 
$\Phi$ be a CQ map.

We now wish to examine in more detail those maps for which
$R_{\Phi}$ is not unitary.  We can assume, without loss of generality,
that the singular values of $R_{\Phi}$  can be written as
$\cos \theta_1$ and $\cos \theta_2$, that
$\cos \theta_1 \geq \cos \theta_2 $, and that $0 \leq \cos \theta_2 < 1$.
Recall that we showed in Lemma 15 of \cite{RSW}
that one can use the singular value decomposition of $R_{\Phi}$ to write
\be
R_{\Phi}  = V \pmatrix{\cos \theta_1 & 0 \cr 0 & \cos \theta_2} W^{\dg}
 = \half U_+ + \half U_-
\ee
 where  $U_{\pm} =
  V \pmatrix{ e^{ i \pm \theta_1} & 0 \cr 0 & e^{ i \pm \theta_2}} W^{\dg}$.
and $V, W$ are unitary.
Thus, $\Phi$ is the midpoint of a line segment in ${\cal S}$
and can be written as
\be \label{eq:convcb2}
  \Phi = \half \Phi_{U+} + \half \Phi_{U-}
\ee
with $ \Phi_{U\pm}$ defined as in (\ref{eq:genU}).  Although
$ \Phi_{U\pm}$ need not have the canonical form (\ref{eq:Tcanon}),
they are related so that their sum does.

We now use the singular value decomposition of $R_{\Phi}$ to
decompose it into unitary maps in another way.
\be
R_{\Phi} &  = &  \label{eq:Rsingval}
   V \pmatrix{\cos \theta_1 & 0 \cr 0 & \cos \theta_2} W^{\dg}  \\
& = &  V \Big[ \frac{\cos \theta_1 + \cos \theta_2}{2} I +
   \frac{\cos \theta_1 - \cos \theta_2}{2} \sigma_z \Big] W^{\dg}
     \nonumber \\   \label{eq:preconvcomb}
& = & \frac{\cos \theta_1 + \cos \theta_2}{2} \, VW^{\dg} +
    \frac{\cos \theta_1 - \cos \theta_2}{2} \, V \sigma_z  W^{\dg} .
\ee
Moreover, it follows from (\ref{eq:preconvcomb}) that
\be   \label{eq:convcb3}
   \Phi = \frac{\cos \theta_1 + \cos \theta_2}{2} \, \Phi_{VW^{\dg}} +
    \frac{\cos \theta_1 - \cos \theta_2}{2} \, \Phi_{V \sigma_z  W^{\dg}}
   + (1 - \cos \theta_1) \Phi_0
\ee
where $\Phi_0$ is the QC map corresponding to
$M = \pmatrix{ \wh{\Phi}(E_{11}) & 0 \cr  0 & \wh{\Phi}(E_{22})}$.
Since we have assumed that we do {\em not} have
$\cos \theta_1 = \cos \theta_2 = 1$, equation (\ref{eq:convcb3})
represents $\Phi$ as a non-trivial convex combination of at
least {\em two} distinct CPT maps, the first two of
which are generalized extreme points.  (Unless
$\cos \theta_1 = 1 $ or $\cos \theta_1 = \cos \theta_2$, we will
have three distinct points, and can already conclude that
$\Phi$ lies in the interior of a segment of a plane within
${\cal S}$.)
Now, the assumption that $\cos \theta_2 \neq 1$ suffices to
show that the decompositions (\ref{eq:convcb3}) and
(\ref{eq:convcb2}) involve different sets of extreme points
and, hence,  that $\Phi$ can
be written as a point on two distinct line segments in ${\cal S}$.
Therefore, there is a segment of a plane in ${\cal S}$ which
contains $\Phi$ and for which $\Phi$ does not lie on the boundary
of the plane (although the plane might be on the boundary of ${\cal S}$).
Thus we have proved the following.
\begin{lemma}
Every map $\Phi$ in ${\cal S}$ lies in one of
two disjoint sets which allows it to be characterized as follows.
Either
\begin{itemize}
\item[I)]  $\Phi$ is a generalized extreme point of ${\cal S}$, or

\item[II)]  $\Phi$ is in the interior of a segment of a
plane in ${\cal S}$.

\end{itemize}
\end{lemma}

Now let ${\cal T}$ denote the set of maps for which
$\Phi \circ T$ or, equivalently $(-I) \circ\Phi$, is in ${\cal S}$.
Since ${\cal T}$ is a convex set  isomorphic to
${\cal S}$, its elements can also be broken into two
classes as above.
The set of entanglement breaking maps is precisely
${\cal S} \cap {\cal T}$.  We can now prove
 Theorem~\ref{sect:convCQ} by showing that 
the convex hull of CQ maps is
  ${\cal S} \cap {\cal T}$.
\pf  Let $\Phi$ be in ${\cal S} \cap {\cal T}$
which is also a convex set.
If $\Phi$ is a generalized extreme point of either
${\cal S}$ or ${\cal T}$, then the only possibility
consistent with $\Phi$ being entanglement-breaking is that
it is CQ.   Thus we suppose that $\Phi$ belongs to class II
for both ${\cal S} $ and $ {\cal T}$.  Then $\Phi$ lies
within a plane in ${\cal S} $ and within a plane in $ {\cal T}$.
The intersection of these two planes is non-empty (since it
contains $\Phi$) and its intersection must contain a line segment
in ${\cal S} \cap {\cal T}$ which contains $\Phi$ and
for which $\Phi$ is {\em not} an endpoint.   Therefore,
$\Phi$ is {\em not} an extreme point of ${\cal S} \cap {\cal T}$.
Thus all possible extreme points of ${\cal S} \cap {\cal T}$ must
be generalized extreme points of ${\cal S}$ or ${\cal T}$, in which
case they are CQ.    ~~~\qed

\smallskip

\rmk Although this shows that all extreme points of  
${\cal S} \cap {\cal T}$ are CQ maps, \linebreak this need not hold for
the various convex subsets, corresponding to allowed values of
$\lambda_k, t_k$ in a fixed basis,
 discussed at the start of Section~\ref{sect:cp}.  The following
remark shows that ``most'' points  in the convex subset
$\Lambda_{\bt, \lambda_3}$ of the $\lambda_1$-$\lambda_2$ plane
can, in fact,  be written as a convex combination of CQ maps in canonical
form in the same basis.  It also shows why it is necessary
to go outside this region for those points close to the boundary.


\begin{itemize}
\item[a)] First consider the set of entanglement-breaking maps
with $\lambda_3 = 0$, which is the convex set
$\cup_{t_3} \Xi_{t_3, 0}$.   Every extreme point
must be an extreme point of the convex set $\Xi_{t_3, 0}$ for some
$t_3$.   By Theorem \ref{thm:RSW}, these are the maps for which
 $\frac{1}{\sqrt{1-t_3^2}} \pmatrix{ \tau  & \lambda_+ \cr
    \lambda_- & \tau}$ is unitary, which implies that either
\begin{itemize}
\item[(i)]  $t_1 = t_2 = 0$ and $(\lambda_1 \pm \lambda_2)^2 = 1 - t_3^2$
which implies that either $\lambda_1 = 0$ or $\lambda_2 = 0$ with
$t_3^2 + \lambda_j^2 = 1$ for $j = 1$ or $2$, {\em or}
\item[(ii)]  $\lambda_1 = \lambda_2 = 0$ and $|\bt|^2 = 1$.
\end{itemize}
The first type of extreme point is obviously a CQ map; the second
is a ``point'' channel which, as noted before, is a special
case of a CQ map.  Thus any map in $\Xi_{t_3, 0}$ can be written
as a convex combination of CQ maps in $\Xi_{t_3, 0}$.

Similar results hold if $\lambda_1 = 0$ or  $\lambda_2 = 0$.
Therefore, any entanglement breaking channel with some $\lambda_k = 0$,
 can be written as a convex combination of CQ channels with at
most one non-zero $\lambda_k$ {\em in the same basis}.
Thus any planar channel can be written as a convex combination
of CQ channels in the same plane.

\item[b)] Next consider entanglement-breaking maps with at most one
non-zero $t_k$.   We can assume, without loss of generality,
 that $t_1 = t_2 = 0$ in which
case the conditions for complete positivity reduce to 
(\ref{eq:cp.tau0}).  Combining this with the sign change condition
  yields
\be
 (|\lambda_1| + | \lambda_2|)^2 \leq (1 - | \lambda_3|)^2 - t_3^2 .
\ee
It follows that for each fixed value of
$\lambda_3$ the set of allowable $(\lambda_1, \lambda_2)$
 form a square with corners  $(0, \pm A_3), (\pm A_3, 0)$ where
$A_3 = \sqrt{ (1 - | \lambda_3|)^2 - t_3^2 }$.  Thus, the extreme
points of $\Lambda_{(0,0,t_3), \lambda_3}$ are planar channels
which, by part(a) are in the convex hull of CQ channels.
In particular, a map with $\lambda_1 = 0, \lambda_2 = \pm A_3$, 
can be written as a convex combination of CQ maps with 
either $\lambda_2 = 0$ or $\lambda_3 = 0$.   However, these
maps need not necessarily lie in $\Lambda_{(0,0,t_3), \lambda_3}$;
we can only be sure that $\lambda_1 = 0$ and $ t_1 = 0$, but
not that $t_2 = 0$.  Thus
we can only state that $\Lambda_{(0,0,t_3), \lambda_3}$ is in
the convex hull of those CQ maps with $\lambda_j = 0$ and $ t_j = 0$
for either $j = 1$ or $2$.   Although it may be necessary to 
enlarge the set $\Lambda_{(0,0,t_3), \lambda_3}$ in order to
ensure that it is in the convex hull of some subset of CQ
maps, these CQ maps will have the canonical form 
{\em in the same basis}, and the same value for $\lambda_3$
in that basis.

\item[c)]  Now consider the convex subset
$\Lambda_{\bt, \lambda_3} \cap \Lambda_{\bt, -\lambda_3}$ of 
the $\lambda_1$-$\lambda_2$ plane corresponding to
entanglement breaking maps with $\bt, |\lambda_3|$ fixed.
These two regions intersect when either $\lambda_1 = 0$ or
$\lambda_2 = 0$ (or, equivalently, $|\lambda_+| = |\lambda_-|$
where $\lambda_{\pm} = \lambda_1 \pm \lambda_2$).   One can
again use part (a) to see that these intersection points can
be written as convex combinations of CQ maps in canonical form
in the {\em same basis}.   Since their convex hull has the same
property, the resulting parallelogram, as shown in 
Figure~\ref{fig:convhull}, is also
a convex combination of CQ maps of the same type.  Only for
those points in the strip between the parallelogram and the
boundary might one need to make a change of basis in order
to write the maps as a convex combination of CQ maps.

\item[d)] Now suppose 
$|\lambda_1| = |\lambda_2| = |\lambda_3| = \lambda > 0 $.
Since any two signs can be changed by conjugation
 with a Pauli matrix, $\Phi$ is unitarily equivalent to a map
 with $\lambda_1 = \lambda_2 = \lambda_3 = \pm \lambda$.
One can then conjugate with another unitary matrix
(corresponding to a rotation on the Bloch sphere) to
conclude that $\Phi$ is unitarily equivalent to a channel $\Phi^{\prime}$
with $t_1 = |\bt| = \sqrt{\sum_k t_k^2 }$, and  $ t_2 = t_3 = 0 $.
It then follows from part (b) that $\Phi^{\prime}$, and thus also
$\Phi$,  can be written as
 a convex combination  of CQ channels which have the
 form described above in the {\em rotated basis}.   However,
these maps need not necessarily have the canonical form in
the original basis.

\end{itemize}

Consider the region $\Lambda_{\bt,\lambda_3}$ with
$0 < \lambda_3 = \lambda< \frac{1}{3}$ and 
$|\bt|^2 = 1 - 2 \lambda  + 3 \lambda ^2$.   The maps with  
$|\lambda_1| = |\lambda_2| = \lambda$ lie on the boundary 
of this region (in fact, at the intersection of the boundary
with the $\lambda_{\pm}$ axes, as shown in
Figure~\ref{fig:convhulloct}).   Since these maps have the form
considered in part (d) they can be written as a convex combination of CQ
maps; however, those CQ maps need not have the canonical form in the
original basis. Nevertheless,  every point in the octagon formed  from
the  convex  hull of  the intersection points of the lines 
$|\lambda_+| = |\lambda_{-}|$, $|\lambda_+| = 0$, and $|\lambda_{-}| = 0$
with the boundary, as shown in 
Figure~\ref{fig:convhulloct}), can be
written as a convex combination of CQ maps as described above. 

As another example, consider the set of entanglement breaking
maps with $\bt = (0,0,t_3)$ fixed.   For any fixed $ \lambda_1$,
the set $\Lambda_{\bt, \lambda_1} \cap
\Lambda_{\bt, -\lambda_1}$ is a convex subset of  the
$\lambda_2$-$\lambda_3$ plane.   Let $(\lambda_2,\lambda_3)$
be a point in this subset that lies between the boundary
and a  parallelogram as described in (c) above.   By considering
the associated map as a point in the set 
$\Lambda_{\bt, \lambda_3} \cap \Lambda_{\bt, -\lambda_3}$ instead, 
one can be sure that it can be written as a convex combination of
CQ maps since this subset of the $\lambda_1$-$\lambda_2$ 
plane is of
the type described in (b).   Moreover, these boundary points
can be added to the convex hull of CQ maps without need for
a change of basis.

One might
expect that additional boundary points could be added in various
ways with additional ingenuity and bases changes.   That this is
always true, is the essence of  Theorem~\ref{thm:CQqubit}.
Only for points near the boundary with two $t_k$ non-zero is
it necessary to actually make the change of basis used in the
proof of this theorem.   In other cases, the necessary convex
combinations (which are not unique) can be formed using the
strategies outlined above.

\bigskip


\noindent{\bf Acknowledgment:} This paper is the outgrowth of
discussions with  a number of people.  I owe a particularly
large debt to Peter Shor for stimulating my interest in this
topic by communicating the results of \cite{Shor} prior to
publication.  Both M. Horodecki and P. Shor
 communicated several other results which 
have now been incorporated into \cite{HRS}.
It is also a pleasure to thank   Elisabeth
Werner for helpful discussions, particularly regarding
section~\ref{sect:convCQ};   Chris King for many helpful
discussions;  John Cortese for
 making stimulating comments, communicating a counter-example to the
sufficiency of the determinant condition,
 and providing many plots, including four of the figures in this paper,
 which helped to clarify my thinking.

\begin{figure}
\includegraphics*[width=6cm,height=6cm,keepaspectratio=true]{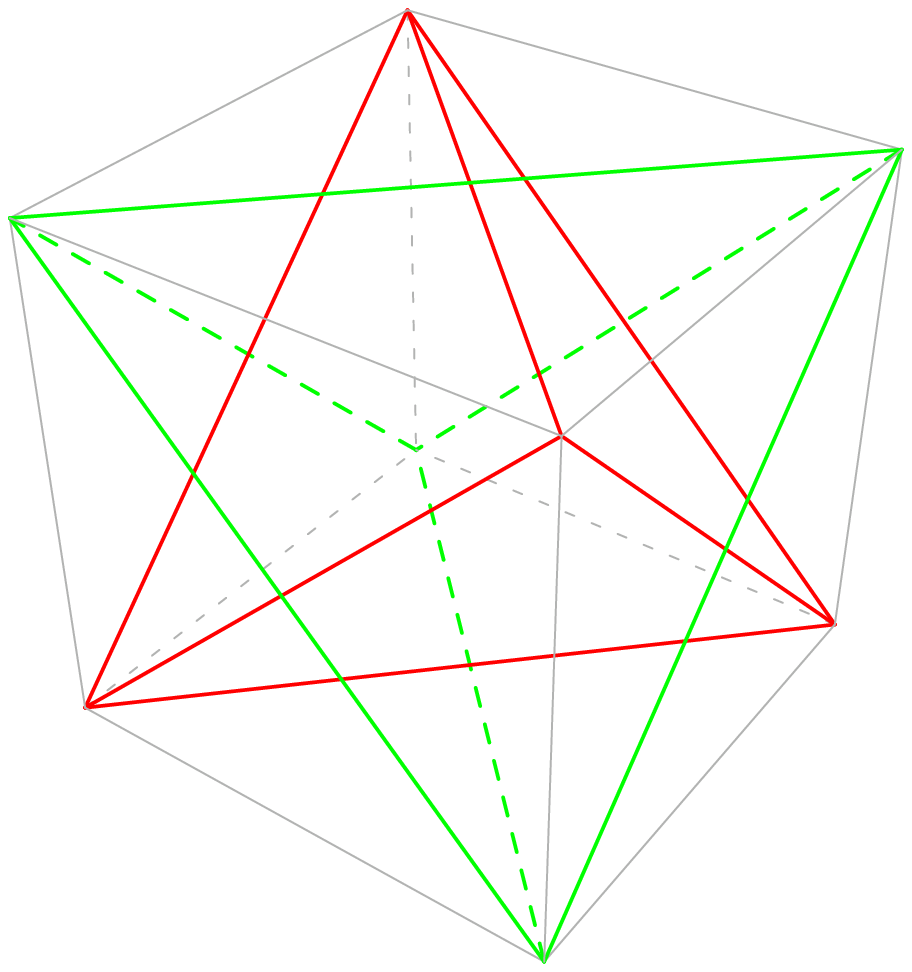}
\hskip2cm
\includegraphics*[width=6cm,height=6cm,keepaspectratio=true]{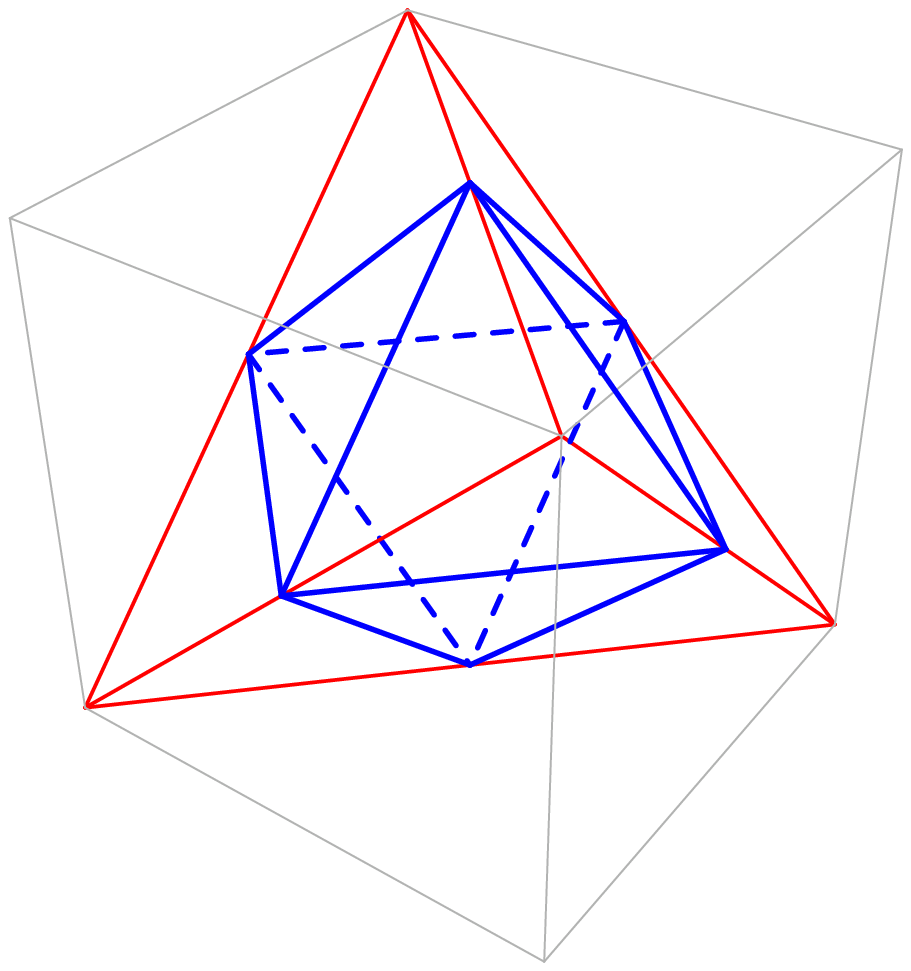}
\caption{The tetrahedron of bistochastic maps and its inversion
through the origin (left);  their intersection
gives the octahedron of unital entanglement breaking maps (right).
(Figures by K. Durstberger appeared in  \cite{BNT}. )}
\label{fig:octa}
\end{figure}

\begin{figure}
\hskip3cm
\includegraphics*[width=7cm,height=7cm,keepaspectratio=true]{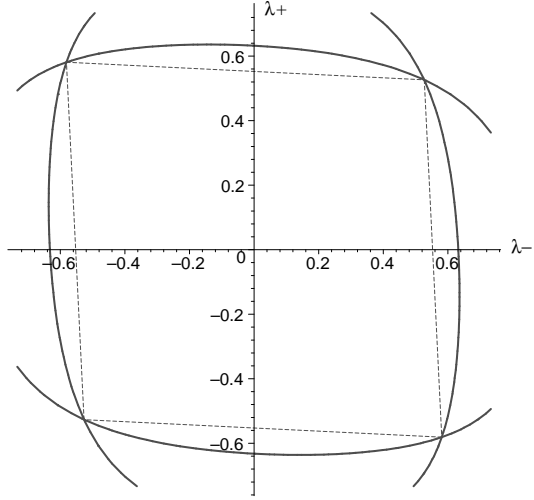}
\caption{The region of the $\lambda_+$-$\lambda_-$ plane corresponding
to entanglement breaking maps with $\bt = (0.4,0.3,0.0)$ and $\lambda_3 =
0.15$.  The dotted lines show the convex hull of the  intersection
points, which are planar maps.}
\label{fig:convhull}
\end{figure}

\begin{figure}
\hskip3cm
\includegraphics*[width=9cm,height=9cm,keepaspectratio=true]{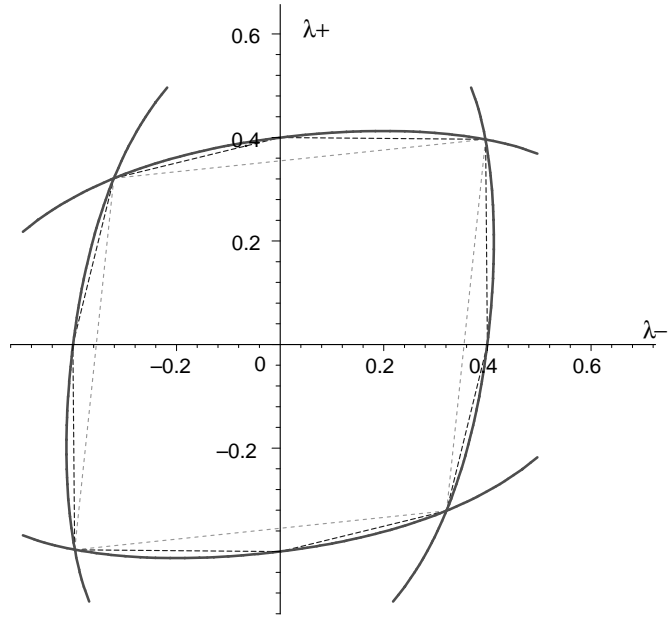}
\caption{The region of the $\lambda_+$-$\lambda_-$ plane corresponding
to entanglement breaking maps with $\bt = (0.4,0.3,0.3742)$ and 
$\lambda_3 = 0.20$,
Because the intersections of the axes with the boundary 
(at $\lambda_{\pm} = \pm 0.4$, for which  all $|\lambda_k| = 0.2)$
correspond to maps known to be in the convex hull of CQ maps,
one can enlarge the convex hull of such maps from the dotted line
to the octagon shown by the dashed line.}
\label{fig:convhulloct}
\end{figure}

Part of this work was done while visiting the
mathematics department at the Amherst campus of the University
of Massachusetts; I am grateful to Professor Donald
St. Mary for arranging this visit and providing a very 
hospitable working environment.  Finally, I would like to thank 
Heide Narnhofer for permission to use the postscript files produced by 
K. Durstberger for Figure~\ref{fig:octa}.

\bigskip




\end{document}